\documentstyle[preprint,prc,aps,psfig]{revtex}
\everymath={\displaystyle}
\begin{document}
\title{\bf On a q-analogue of the spin-orbit coupling}

\author{M. Micu}
\address{Department of Theoretical Physics, Horia Hulubei Institute of
Physics and Nuclear Engineering, POB MG-6, Bucharest, 76900 Romania}
\author{Fl. Stancu}
\address{University of Li\`ege, Institute of Physics B5, Sart Tilman, B-4000
Li\`ege 1, Belgium}

\date{\today}
\maketitle

\begin{abstract}
\baselineskip=0.50cm
Based on the tensor method, a $q$-analogue of the spin-orbit coupling is
introduced in a $q$-deformed
Schr\" odinger equation, previously derived for a central potential.
Analytic expressions for the matrix elements of the representations
$j = \ell \pm 1/2$ are derived. The spectra
of the harmonic oscillator and the Coulomb potential are calculated numerically
as a function of the deformation parameter, without and with the spin-orbit
coupling. The harmonic oscillator spectrum presents strong analogies with the
bound spectrum of an Woods-Saxon potential customarily used in nuclear physics.
The Coulomb spectrum simulates relativistic effects. The addition of the 
spin-orbit coupling reinforces this picture.
\end{abstract}
\section{Introduction}
A particular interest has been devoted during the last decade to the
quantum algebra $su_q(2)$ \cite{KU81,SK82,JI86,BI89,MA89}. This algebra is
generated by three operators $L_+,~L_0$ and $L_-$, also named the $q$-angular
momentum components. They have the following commutation relations:
\begin{equation}\label{DEF1}
\left[~L_0~,~L_\pm~\right]~=~\pm~L_\pm,
\end{equation}
\begin{equation}\label{DEF2}
\left[~L_+~,~L_-~\right]~=~\left[2~L_0\right],
\end{equation}
where the quantity in square brackets is defined as
\begin{equation}
$$\left[n\right]~=~{q^n-q^{-n}\over q-q^{-1}}.
\end{equation}
 
In the most general
case the deformation parameter $q$ is an arbitrary complex number and the
physicist considers it as a phenomenological parameter \cite{BO91}. When $q=1$,
the quantum algebra
$su_q(2)$, which defines a $q$-analogue of the angular momentum, reduces to the
Lie algebra $su(2)$ of the ordinary angular momentum.\par
It is therefore interesting to investigate $q$-analogues of dynamical systems
and to look for new effects when $q \ne 1$. This has been first achieved for
the harmonic oscillator by using algebraic methods, as e.g. in Refs.
\cite{BI89,MA89}. Taking, for example, $q = \exp(is)$ with $s$ a real,
positive quantity, one can find that the distance between subsequent levels of
the $q$-harmonic oscillator decreases when the excitation
increases. This is a desired property in describing rotational bands of
deformed nuclei \cite{BO91}. However the accidental degeneracy of the harmonic
oscillator persists in this treatment.\par
Another, more appealing way to introduce $q$-analogues of simple dynamical
systems, is through deriving a $q$-deformed Schr\"odinger equation. In this
vein several attempts have been made for the 
harmonic oscillator, as for example in Refs.\cite{MINAHAN,LI92,CAROW},  
for an attractive Coulomb potential 
\cite{XI92,IR66} or for both potentials \cite{PAPP,MI99}. This procedure 
leads to the removal of the accidental degeneracy 
whenever it exists. \par
Here we follow the approach of Ref. \cite{MI99} where a $q$-deformed
Schr\"odinger equation has been derived for a general central potential and the
exact solution for the particular cases of the Coulomb and the harmonic
oscillator potentials have been obtained. The crucial starting point in Ref.
\cite{MI99} was the search for a hermitian realization of the position,
momentum
and angular momentum operators, all behaving as {\it{vectors}} with respect to
$su_q(2)$ algebra. This allowed the construction of an angular momentum
operator entering the expression of the Hamiltonian. Its components are
different from the generators of the $su_q(2)$ algebra. In the case of central
potentials (spinless particles) the eigenfunctions of the $q$-deformed angular
momentum have been derived as $q$-deformed spherical harmonics and then closed
expressions for the eigenvalues of the $q$-deformed Schr\"odinger equation have
obtained as a function of $q$.\par
This study is   
devoted to the derivation of a $q$-deformed
spin-orbit coupling, consistent with the approach of Ref. \cite{MI99}.
There an angular momentum  $\Lambda_{\mu}$ ($\mu = 0, \pm1$)
has been defined as a $q$-{\it vector} with respect to the $su_q(2)$
algebra (\ref{DEF1})-(\ref{DEF2}). By analogy, here we introduce a spin
operator $\sigma_\mu$ which is a $q$-{\it vector} in the algebra
of $S_\mu$ analogue to (\ref{DEF1})-(\ref{DEF2}). Next a hermitian spin 
operator $\Sigma_{\mu}$ is constructed as a $q$-{\it vector} in the coproduct
algebra of the total angular momentum $J_\mu$ 
by using a unitary version of the universal $R$ matrix.
The spin-orbit interaction is defined as a $q$-scalar hermitian operator
in the space of $J_\mu$ and its matrix elements are calculated
exactly for the representations $j = \ell \pm 1/2$.\par

In previous applications of the $q$-deformed algebras to physical systems,
as for example Ref. \cite{RAYCHEV98}, the spin-orbit
coupling is derived in a different way, based on a boson
realization of the $so_q(3)$ algebra \cite{RAYCHEV96}.
There the spin operator 
does not form a vector in the coproduct algebra.  
Accordingly
the eigenvalues of the spin-orbit operator
are different from ours.\par

In the next section we summarize the findings of Ref. \cite{MI99}.
In Section III a $q$-analogue of the spin-orbit coupling is
derived. In Section IV we calculate numerically the spectra
of the $q$-harmonic oscillator and the $q$-Coulomb
potentials without and with a spin-orbit contribution. 
Physical implications are discussed. We stress that we do
not aim at a particular fit of the deformation parameter to describe some
particular system but at modelling physical systems through $su_q(2)$ algebra.
The last section is devoted to some closing remarks.
\section{Spinless particles}
In this section we follow closely Ref. \cite{MI99}. The Hamiltonian entering
the $q$-deformed Schr\"odinger equation is
\begin{equation}
{\sl{H}} = \frac{1}{2}~\vec{p}\ ^2 + V(r)~.
\end{equation}
Here and in the following we shall take 
\begin{equation}
\hbar = c = e = m  = 1~.
\end{equation}
The eigenfunctions of this Hamiltonian are
\begin{equation}\label{YLM}
\Psi(r,x_0,\varphi) = r^L u_L(r) Y_{\ell m}(q,x_0,\varphi)~,
\end{equation}
where $Y_{\ell m}(q,x_0,\varphi)$ are the normalized $q$-spherical harmonics
(56) and (57) of Ref. \cite{MI99},
depending of the deformation parameter $q$ and $x_0 = cos \ \theta$.
They are related to $q$-hypergeometric functions \cite{ANDREWS}
.\par
The function $r^Lu_L(r)$ satisfies the following radial equation
\begin{equation}
\left\{\frac{1}{2} \left[- \left(\frac{\partial^2}{\partial r^2} + \frac{2}{r}
\frac{\partial}{\partial r}\right) + \frac{1}{r^2} L(L+1)\right] +
V_0(r)\right\} r^Lu_L(r) = E_{n\ell}~r^Lu_L(r)~,
\end{equation}
where $L$ is the non-negative solution of
\begin{equation}\label{LLPLUS1}
L(L+1) = \frac{\left[2 \ell\right]}{\left[2\right]} \frac{\left[2 \ell
+2\right]}{\left[2\right]} + c_{\ell}^2 - c_{\ell}~,
\end{equation}
with
\begin{equation}\label{CL}
c_{\ell} = \frac{q^{2 \ell +1} + q^{-2 \ell -1}}{\left[2\right]}~.
\end{equation}
It then follows that for the Coulomb potential
\begin{equation}
V_0(r) = - \frac{1}{r}~,
\end{equation}
the eigenvalue is
\begin{equation}
\left(E_{n \ell}\right)_{Coulomb} = - \frac{1}{2(n+L+1)^2}~,
\label{COUL}
\end{equation}
and for the harmonic oscillator potential
\begin{equation}
V_0(r) = \frac{1}{2}~r^2~,
\end{equation}
the eigenvalue is
\begin{equation}
\left(E_{n \ell}\right)_{oscillator} = (2n+L+ \frac{3}{2})~.
\label{HO}
\end{equation}
$n$ being in both cases the radial quantum number.\par
The spectrum is degenerate with respect to the magnetic quantum number $m$ but
the accidental degeneracy typical for the undeformed equation is removed both
for the Coulomb and the harmonic oscillator potentials when $q \ne 1$.\par
From Eq. (\ref{CL}) it follows that for $\ell =0$ one has $c_{\ell}=1$.
Thus for $\ell =0$ the only non-negative solution of 
(\ref{LLPLUS1}) is $L=0$, for all
deformations. As a consequence, the $\ell =0$ levels are independent of the
deformation parameter both for the harmonic oscillator and the Coulomb
potentials. The centrifugal barrier disappears and taking $V_0(r)=0$ one
reobtains the free particle case, as for undeformed equations.\par
For $\ell \ne 0$ it is useful to distinguish between two different types of
deformation parameter:

\medskip

\begin{equation}
(i)\,\,\,\, q=e^{s} \,\,\,\,\,\mbox{with} \  s \ \mbox{real}~.
\label{QREAL}
\end{equation}
In this case one can easily prove that $c_{\ell} \ge 1$ so that Eq.
(\ref{LLPLUS1}) has
real solutions. Therefore to each non-zero $\ell$ corresponds a positive $L$
which is no more an integer. We found it
interesting to use real $q$ for the Coulomb potential, as shown in Sec. III.
The other case is:
\begin{equation}
(ii)\,\,\,\,q=e^{is} \,\,\,\,\,\mbox{with}\  s \ \mbox{real}~.
\label{QCOM}
\end{equation}
In this case for small values of $s$ one can find numerically 
that  real positive values of $L$ exist. This
case is applicable to the harmonic oscillator potential, because it leads to
interesting analogies of its spectrum with a known case in nuclear physics,
as discussed in Sec. III
\par
\section{Derivation of the spin-orbit coupling}
Now the Hamiltonian (4) contains a potential of the form
\begin{equation}
V = V_0(r) + \alpha(r) V_{S-O}~,
\end{equation}
where $V_0$ is the central potential from the previous section, $V_{S-O}$ the
spin-orbit operator and $\alpha$ a function which vanishes when
$r \rightarrow \infty$. In atomic or nuclear physics the spin-orbit
operator is the ordinary scalar product between the spin and angular
momentum. In the deformed case considered here we aim at introducing
a similar definition. However there are inherent differences due to
the more complex nature of the q-deformed vector operators, as
explained below.\par

By analogy to the q-angular momentum $L_\mu$
one can define a spin operator $S_\mu$ through relations similar to (1-3).
The operators $L_\mu$ and $S_\mu$
satisfy the hermiticity relations
\begin{eqnarray}\label{r2}
&&L^\dagger_{\pm,0}=L_{\mp,0}\nonumber\\
&&S^\dagger_{\pm.0}=S_{\mp,0}
\end{eqnarray}
However the situation is different from the $su(2)$ case because
neither $L_\mu$ nor $S_\mu$ form a vector with respect 
to their $su_q(2)$ algebra.
In a $su_q(2)$ algebra a q-vector of components $V_i~ (i=0,\pm 1)$,
is defined through the relations \cite{MI99}
\begin{eqnarray}\label{r4}
&&(L_\pm V_i-q^i V_i L_\pm)q^{L_0}=\sqrt{[2]}V_{i\pm1}\nonumber\\
&&[L_0,V_i]=iV_i
\end{eqnarray}
where one takes $V_{\pm2}$ = 0 whenever it appears.

But as in 
Ref. \cite{MI99}, instead of $L_\mu$ we have to 
use $\Lambda_\mu$ defined as
\begin{equation}\label{LAPM}
$$\Lambda_{\pm1}~=~\mp ~\sqrt{\frac{1}{[2]}}~q^{-L_0}~L_\pm,
\end{equation}
\begin{equation}\label{LAZERO}
\Lambda_0~=~{1\over[2]}~\left(q~L_+~L_-~-~q^{-1}~L_-~L_+\right).
\end{equation}
These quantities form a vector in the $su_q(2)$ algebra, i.e. satisfy 
relations the (\ref{r4}) as it can be easily checked. 
By analogy to (\ref{LAPM}) and (\ref{LAZERO}) 
we introduce a vector of components $\sigma_\mu$ in the $su_q(2)$
algebra having $S_\mu$ as generators
\begin{equation}\label{SIPM}
$$\sigma_{\pm1}~=~\mp ~\sqrt{\frac{1}{[2]}}~q^{-S_0}~S_\pm,
\end{equation}
\begin{equation}\label{SIZERO}
\sigma_0~=~{1\over[2]}~\left(q~S_+~S_-~-~q^{-1}~S_-~S_+\right).
\end{equation}
\par
In the space generated by $S_\mu$ the quantities $L_\mu$ are scalars
and vice versa, which implies that 
\begin{equation}\label{COMM}
[ \sigma_{\mu},\Lambda_{\mu'} ] = 0.
\end{equation}
In dealing with the spin-orbit operator we have to also introduce
the coproduct algebra of $L_\mu$ and $S_\mu$. The generators $J_{\mu}$ 
of this algebra are defined as 
\begin{equation}\label{JPM}
J_\pm=L_\pm q^{-S_0}+S_\pm q^{L_0},
\end{equation}
\begin{equation}\label{JZERO}
J_0=L_0 + S_0~.
\end{equation}
One can directly prove that they satisfy commutation relations 
of type (\ref{DEF1}) and (\ref{DEF2}). 
One can also prove that $\Lambda_\mu$ are the 
components of a vector in the coproduct algebra
which means that they satisfy relations analogous to (\ref{r4}) 
with $J_{\mu}$ instead of $L_{\mu}$.
On the other hand $\sigma_\mu$ do not fulfil such relations.
However, instead of $\sigma_\mu$ one can introduce another vector $\Sigma_\mu$
satisfying relations of type (\ref{r4}) with $J_{\mu}$ instead of $L_{\mu}$.
This can be achieved by using the universal $R$ matrix. 
In fact
we need both the $R$ matrix and its conjugate \cite{CU91}.
The latter is here denoted by $\mathcal R$.\par

The $R$ matrix or its conjugate has the property that it
replaces $q$ by $q^{-1}$
in definition (\ref{JPM}) i.e. one has
\begin{equation}\label{r8}
R\left(L_\pm~q^{-S_0}~+~S_\pm~q^{L_0}\right)~=~\left(L_\pm~q^{S_0}~+~
S_\pm~q^{-L_0}\right)~R,
\end{equation}
and
\begin{equation}\label{r8prim}
{\mathcal R}\left(L_\pm~q^{-S_0}~+~S_\pm~q^{L_0}\right)~=~\left(L_\pm~q^{S_0}~+~
S_\pm~q^{-L_0}\right)~{\mathcal R}.
\end{equation}
The operator (\ref{JZERO}) remains unchanged or in other words
\begin{equation}\label{COM}
[R,~J_0] = 0, \,\,\,\,\ [{\mathcal R},~J_0] = 0.
\end{equation}
We found it convenient to use the $R$ matrix as defined in Ref. \cite{RS92}.
For $s=1/2$ it contains two terms only
\begin{equation}\label{RQ}
R=q^{2L_0S_0}+{\lambda\over\sqrt{q}}L_-S_+~,
\end{equation}
where
\begin{equation}
\lambda = q - 1/q ~. 
\end{equation}
One can check that the expression (\ref{RQ}) satisfies (\ref{r8}).
The conjugate $\mathcal R$ of $R$ takes the form
\begin{equation}\label{ROVERQ}
{\mathcal R}=q^{-2L_0S_0}-\lambda\sqrt{q}L_+S_-~,
\end{equation}
and it satisfies equation  (\ref{r8prim}).
Using (\ref{RQ}) and (\ref{ROVERQ}) one defines \cite{QUESNE}
\begin{equation}\label{SIGQ}
\Sigma_\mu(R)=R^{-1}\sigma_\mu R~,
\end{equation}
and
\begin{equation}\label{SIGOVERQ}
\Sigma_\mu({\mathcal R})={\mathcal R}^{-1}\sigma_\mu{\mathcal R}~.
\end{equation}
The operator (\ref{SIGQ}) with $\mu = 0,\pm1$ 
form a vector in the space of the coproduct algebra. The proof is given
in Appendix A. In a similar way one can prove that the operator 
(\ref{SIGOVERQ}) is also a vector in the coproduct algebra.

Note that none of the above operators
is hermitian but each $\mu$-component of one is related to the
corresponding component of the other through the relation 
\begin{equation}\label{HERM}
\Sigma_\mu^+(R)=\left(-{1\over q}\right)^\mu \Sigma_{-\mu}({\mathcal R})~.
\end{equation}
relating operators associated with $R$ and $\mathcal R$.
To overcome the lack of hermiticity one can make use of the unitary matrix
$R_u$ introduced in Ref. \cite{CU91} as
\begin{equation}\label{r17}
R_u={1\over N}\left(\sqrt{q} R+{1\over\sqrt{q}}{\mathcal R}\right)
\end{equation}
where $N=q^{l+1/2}+q^{-l-1/2}$ is a normalization factor.
With the help of $R_u$ one can define the vector 
\begin{equation}\label{r15}
\Sigma_\mu(R_u)=R_u^\dagger~\sigma_\mu~R_u
\end{equation}
the components of which are hermitian operators, i.e. satisfy the relation
\begin{equation}\label{r16}
\Sigma_\mu^\dagger(R_u)=\left(-{1\over q}\right)^\mu
\Sigma_{-\mu}(R_u).
\end{equation}
Now we can define a hermitian spin-orbit operator as
\begin{equation}\label{r18}
V_{S-O}={1\over2} \vec{\Sigma}(R_u)\vec{\Lambda}+
{1\over2} \vec{\Lambda}\vec{\Sigma}(R_u)~,
\end{equation}
where the scalar product between the $q$-vectors $ \vec{\Sigma}(R_u)$
and $\vec{\Lambda}$ is defined as in Ref. \cite{MI99}
\begin{equation}\label{r20}
\vec{\Sigma}(R_u)\vec{\Lambda}=~\left(-{1\over q}
\right)^\mu\Sigma_\mu (R_u)\Lambda_{-\mu}.
\end{equation}
with an implied summation over $\mu$.
Using (\ref{r15}) one can rewrite (\ref{r18}) as
\begin{eqnarray}\label{r19}
V_{S-O} &=&{1\over 2}
(R^+_u\vec{\sigma}R_u\vec{\Lambda}+hermitian~ conjugate)\nonumber\\
&=&{1\over2}(R^+_u\vec{\sigma}R_u\vec{\Lambda}+
\vec{\Lambda}R^+_u\vec{\sigma}R_u).
\end{eqnarray}
Let us consider the first term in the right hand side of (\ref{r19})
where $R_u$ is replaced by its definition (\ref{r17})
\begin{equation}\label{r22}
{1\over 2}R^+_u\vec{\sigma}R_u\vec{\Lambda}={1\over 2N^2}
~\left(\sqrt{q} R^++{1\over\sqrt{q}}{\mathcal
R}^+\right)\vec{\sigma}\left(\sqrt{q}~R+{1\over\sqrt{q}}{\mathcal
R}\right)\vec{\Lambda}.
\end{equation}
Here we look for example at the term
$(\sqrt{q}R^++{1\over\sqrt{q}}{\mathcal
R}^+)\vec{\sigma}\sqrt{q} R$ 
which can be rewritten by inserting the identity
$RR^{-1} = 1$
in front of $\vec{\sigma}$ and also using the property
${\mathcal R}^+ R = 1.$
This gives
\begin{eqnarray}
(\sqrt{q}R^++{1\over\sqrt{q}}{\mathcal
R}^+)\vec{\sigma}\sqrt{q} R \vec{\Lambda} = 
q~R^+R~\vec{\Sigma}(R)\vec{\Lambda}~+~\vec{\Sigma}(R)\vec{\Lambda}~
=~(1+q~R^+R)\vec{\Sigma}(R)\vec{\Lambda}.
\end{eqnarray}
In a similar way the other term of (\ref{r22}) becomes 
\begin{eqnarray}\label{r25}
(\sqrt{q}R^++{1\over\sqrt{q}}{\mathcal R}^+)\vec{\sigma}
{1\over\sqrt{q}}{\mathcal R}\vec{\Lambda} =
(1+{1\over{q}} {\mathcal R}^+{\mathcal R})\vec{\Sigma}({\mathcal R})\vec{\Lambda}
\end{eqnarray}
where we have used ${\mathcal R~R}^{-1} = 1$ and
$R^+{\mathcal R} = 1.$
Thus
\begin{equation}
{1\over 2}R^+_u\vec{\sigma}R_u\vec{\Lambda}=
{1\over2N^2}\left[(1+q~R^+R)~
\vec{\Sigma}(R)\vec{\Lambda}
~+~(1+{1\over q}
{\mathcal R}^+{\mathcal R})~\vec{\Sigma}({\mathcal R})\vec{\Lambda}\right].
\end{equation}
One can see that in the above relation the vectors $\vec{\Sigma}$ and
$\vec{\Lambda}$ are next to each other as it should be in a $q$-scalar
product.
For the second term of (\ref{r19}) we have
\begin{equation}\label{r27}
{1\over 2} \vec{\Lambda}~R_u^{+}~\vec{\sigma}~R_u = {1\over 2N^2}~
~\vec{\Lambda}~(\sqrt{q}~R^+~+~{1\over\sqrt{q}}~{\mathcal
R}^+)~\vec{\sigma}~(\sqrt{q}~R~+~{1\over\sqrt{q}}~{\mathcal R})~,
\end{equation}
or using
\begin{eqnarray}\label{r28}
&&{\mathcal R}^+=R^{-1}\nonumber\\
&&R^+={\mathcal R}^{-1}
\end{eqnarray}
in the manner explained above, we get
\begin{equation}\label{r29}
{1\over 2} \vec{\Lambda}~R_u^{+}~\vec{\sigma}~R_u =
{1\over2N^2}\left[\vec{\Lambda}\vec{\Sigma}({\mathcal
R})~(1+q~R^+R)~+~\vec{\Lambda}\vec{\Sigma}(R)~(1+{1\over q}
{\mathcal R}^+{\mathcal R})\right].
\end{equation}
Thus the spin-orbit interaction 
takes the form:
\begin{eqnarray}\label{OPERATOR}
V_{S-O}={1\over2N^2}
[(1+q~R^+R)~\vec{\Sigma}(R)\vec{\Lambda}
~+~(1+{1\over q}
{\mathcal R}^+{\mathcal R})~\vec{\Sigma}({\mathcal R})\vec{\Lambda}
\nonumber\\
+\vec{\Lambda}\vec{\Sigma}({\mathcal R})~(1+q~R^+R)~
+~\vec{\Lambda}\vec{\Sigma}(R)~(1+{1\over q}
{\mathcal R}^+{\mathcal R})]
\end{eqnarray}
i.e. it contains the operators:
\begin{equation}\label{SCALARS}
\vec{\Sigma}(R)\vec{\Lambda},~\vec{\Sigma}({\mathcal 
R})\vec{\Lambda},~\vec{\Lambda}\vec{\Sigma}(R),
~\vec{\Lambda}\vec{\Sigma}({\mathcal R}),
~R^+R,~{\mathcal R}^+{\mathcal R}.
\end{equation}
These are scalars because they commute with
$J_i ( i=0,\pm 1)$. In particular, for the last two operators,
the commutation with $J_0$ follows directly from
(\ref{COM}). To prove the commutation with $J_{\pm}$ 
we have to use Eqs. (\ref{r8}) and (\ref{r8prim}).
For example in the case of $R^+R$ we have  
\begin{eqnarray}\label{r8com}
R^+R\left(L_\pm~q^{-S_0}~+~S_\pm~q^{L_0}\right)~
&=&R^+~\left(L_\pm~q^{S_0}~+~
S_\pm~q^{-L_0}\right)~R\nonumber\\
&=&R^+~\left(L_\pm~q^{S_0}~+~S_\pm~q^{-L_0}\right)~{\mathcal R}~R^+R\nonumber\\
&=&~R^+{\mathcal R}~
\left(L_\pm~q^{-S_0}~+~S_\pm~q^{L_0}\right) ~R^+R\nonumber\\
&=&\left(L_\pm~q^{-S_0}~+~S_\pm~q^{L_0}\right) ~R^+R
\end{eqnarray}
where after the second equality sign alternative forms of eqs. (\ref{r28})
have been used.

We can obtain the expectation value of  $V_{S-O}$ for states
of total angular momentum $j=\ell \pm 1/2$ by calculating 
the expectation values of the scalars (\ref{SCALARS}). The simplest way is to 
use the state of maximum weight with $m = j$.  
For $j=l+1/2$ this state reads
\begin{equation}
\psi_{\ell+1/2,\ell+1/2}=Y_{\ell \ell}~\chi_{1/2}~,
\end{equation}
where $Y_{\ell m}$ are defined by Eqs. (56) and (57) of Ref. \cite{MI99} 
and $\chi_{m_s}$ with $m_s = \pm1/2$ is the $s = 1/2$ spinor.
In this case one can show that
the last two operators of the list (\ref{SCALARS})
have the following expectation values:
\begin{eqnarray}\label{RCRUCERP}
&&\langle R^+R \rangle_{\ell+1/2} =q^{2l}\nonumber\\
&&\langle {\mathcal R}^+{\mathcal R} \rangle_{\ell+1/2} =q^{-2l}
\end{eqnarray}
For $j=\ell - 1/2$ and $m = j$ the 
wave function is:
\begin{equation}
\psi_{\ell-1/2,\ell-1/2}={1\over\sqrt{[2\ell+1]}}\left(\sqrt{[2\ell]\over
q}~Y_{\ell \ell}\chi_{-1/2}-q^l~Y_{\ell, \ell-1}\chi_{1/2}\right)~.
\end{equation}
In this case 
the last two operators of (\ref{SCALARS})
have the following expectation values:
\begin{eqnarray}\label{RCRUCER}
&&\langle R^+R \rangle_{\ell-1/2} =q^{-2l-2}\nonumber\\
&&\langle {\mathcal R}^+{\mathcal R} \rangle_{\ell-1/2} =q^{2l+2}
\end{eqnarray}
Both for $j=\ell+1/2$ and $j=\ell-1/2$ 
the proof is similar to that given in the Appendix B for the other 
scalars of (\ref{SCALARS}). Using all these expectation values 
in the case where $j=\ell+1/2$ one can easily
show that the expectation value of $V_{S-O}$ is:
\begin{equation}\label{EPLUS}
E_{\ell+1/2} =
\frac{\left[2\ell\right]}{\left[2\right]^2}~
\frac{q^{l+5/2}+q^{-l-5/2}}{q^{l+1/2}+q^{-l-1/2}}~.
\end{equation}
In a similar but somewhat longer way the following
expectation value of $V_{S-O}$ is obtained for $j=\ell - 1/2$:
\begin{equation}\label{EMINUS}
E_{\ell-1/2} =
-{[2l+2]\over[2]^2}~{q^{l-3/2}+q^{-l+3/2}\over
q^{l+1/2}+q^{-l-1/2}}
\end{equation}
The proof of Eqs. (\ref{EPLUS}) and (\ref{EMINUS}) is given in
Appendix B.
In the limit $q \rightarrow 1$ $E_{\ell+1/2}$ and $E_{\ell-1/2}$  
recover the expectation values of the 
non-deformed spin-orbit coupling $\vec{s} \cdot \vec{\ell}$
namely $\ell/2$ for $j = \ell + \frac{1}{2}$ and $-(\ell + 1)/2$ for $j = \ell -
\frac{1}{2}$ respectively.\par
\section{Numerical results}
In Fig. 1 we represent the eigenvalues (\ref{COUL})
of the Coulomb potential as a
function of $s$ (real), when $q=e^{s}$ ( Eq. (\ref{QREAL})).
One can see that every $E_{n \ell}$
increases with $s$ when $\ell \ne 0$, the reason being that one has $L >
\ell$ when one chooses $q$ to be real. Therefore at a given $q \ne 1$ one has
\begin{equation}\label{KG}
E_{2p} > E_{2s} ~;~~ E_{3d} > E_{3p} > E_{3s}~;~  \mbox{etc.}
\end{equation}
These inequalities are similar to those satisfied by the eigenvalues of the
Klein-Gordon equation for which one has
$E(n,\ell) < E(n-1,\ell+1)$ for fixed $n+\ell+1$
\cite{GR91,GR94}. One expects similar inequaltities to be also satisfied by
the eigenvalues of the spinless
Bethe-Salpeter (or Herbst) equation for a particle in
an attractive Coulomb potential \cite{LU97}.
In fact as long as $Z \alpha < \pi/2$ where $Z$ is the charge
and $\alpha$ the fine structure constant the expansion of the eigenvalues
of the Herbst equation coincides with that of the Klein-Gordon equation \cite{MA}.
Thus the results shown in Fig. 1 suggest that the splitting found for $q \ne 1$
can simulate relativistic kinematic effect.\par
In Fig. 2 the eigenvalues (\ref{HO}) of the harmonic oscillator potential are
plotted as a function of $s$, where $s$ and the deformation parameter are
related by Eq. (\ref{QCOM}).
This choice is based on the fact that it implies $L <
\ell$ so that in the interval $0 < s < 0.13$ the $q$-deformed spectrum
satisfies inequalities as
\begin{equation}\label{WOODS}
E_{1d} < E_{2s}~;~ E_{1f} < E_{2p}~;~E_{1g} < E_{2d}~;~ \mbox{etc.}
\end{equation}
which correspond to a potential the form of which is
between a harmonic oscillator and a square well potential. In nuclear physics
\cite{BO69} the standard form is the Woods-Saxon potential
\begin{equation}\label{WS}
\begin{array}{c}
V(r) = Vf(r)~,\\
f(r) = \left[1 + exp \left(\frac{r-R_0}{a}\right)\right]^{-1}~,
\end{array}
\end{equation}
depending on three parameters $V$, $R_0$ and $a$. In the limit $a
\rightarrow 0$ one approaches
a square well potential of radius $R_0$ and depth $V$. The bound
spectrum of a potential of type (\ref{WS})
satisfies the inequalities (\ref{WOODS}) (see Figs. 2-23
of Ref. \cite{BO69}).\par
Next we add the spin-orbit contribution. 
To single out the role of $V_{S-O}$ here we choose $\alpha$ to be a constant. 
In Fig. 3 we plot $\left(E_{n \ell}\right)_{Coulomb} + \alpha E_{\ell \pm 1/2}$
as a function of $s$, where $s$ is related to $q$ by Eq. (\ref{QREAL}). The
levels are labelled by $n \ell j$ where $\ell$ is the value of the angular
momentum at $q=1$ and $j= \ell \pm \frac{1}{2}$. With $\alpha > 0$ one always
has $j = \ell + \frac{1}{2}$ levels above the $j = \ell - \frac{1}{2}$ levels
due to Eqs. (\ref{EPLUS}) and (\ref{EMINUS}). For convenience we choose $\alpha$ = 0.001. 
We therefore see that the energies increase with increasing $j$ for fixed $\ell$
and increasing $n$ or $\ell$ for fixed $j$. Such a pattern corresponds to
solutions of the Dirac equation for a Coulomb potential 
plus a perturbation which
removes the two-fold degeneracy of the eigenvalues for a Coulomb field.
In Ref.
\cite{GR92} it has been shown that for a Dirac particle moving in a purely
attractive potential the level sequence is
\begin{equation}
2p_{3/2} > 2p_{1/2} > 2s_{1/2}, 
\end{equation}
\begin{equation}
3d_{5/2} > 3d_{3/2} > 3p_{3/2} > 3p_{1/2} >
3s_{1/2},~ \mbox{etc.}
\end{equation}
which is here the case when $s > 0.11$ for the first and when $s > 0.17$  
for both rows of inequalities respectively.
Such sequences are expected for alkaline
atoms.\par
In a similar way we add the spin-orbit coupling (\ref{EPLUS}) 
and (\ref{EMINUS}) 
to $(E_{n \ell})_{oscillator}$ of Eq. (\ref{HO}) and in Fig. 4 we plot
$(E_{n \ell})_{oscillator} + \alpha E_{\ell \pm 1/2}$ as a function of $s$,
where
$s$ is related to $q$ via Eq. (\ref{QCOM}). For the sake of 
the discussion here we choose $\alpha$ = - 0.1
The addition of a spin-orbit coupling
to $(E_{n \ell})_{oscillator}$ brings us a picture even closer to the single
particle spectra encountered in nuclear physics. 
Provided $\alpha$ is negative the level sequence of Fig. 4
is similar to that of the neutron single particle spectrum (see e.g. Figs. 2-30
of Ref. \cite{BO69}).
Also Hartree-Fock calculations based on effective density dependent nucleon-nucleon
interactions \cite{BR70} give a similar spectrum.\par
\section{Summary}
We have constructed a $q$-analogue of the spin-orbit coupling for being
used in a $q$-deformed Schr\"odinger equation previously derived
for a central potential.
The spin-orbit coupling is a $q$-scalar product between
the angular momentum  $\Lambda_\mu$ and the spin operator
$\Sigma_\mu$ both defined a $q$-vectors in the coproduct algebra of
the generators $J_\mu$. The spin operator
has been obtained with the help of a hermitian
form of the universal $R$ matrix. Accordingly, our result is new and
entirely different from previous work 
on the spin-orbit coupling.\par
Numerically
we have shown that the $q$-deformed Schr\"odinger equation for a spinless
particle in a Coulomb field has a spectrum which simulates relativistic
effects. The removal of the accidental degeneracy by a real deformation of type
$q = e^{s}$ with $s > 0$ leads to a level sequence similar to that of the
Klein-Gordon or of the Herbst equations. With the addition of a spin-orbit
coupling the level sequence is close to that of alkaline atoms.\par
The $q$-deformed Schr\"odinger equation for a spinless particle in a harmonic
oscillator potential has a spectrum similar to that of the bound spectrum of an
Woods-Saxon potential. The deformation is of type $q = e^{is}$ with $s$
real and positive.
The addition of a spin-orbit coupling leads to a spectrum
similar to single particle spectra of nuclei.
It would be interesting to pursue this study in a more quantitative way. \par
\section{Appendix A}
In this Appendix we prove that the operators 
\begin{equation}\label{A1}
\Sigma_\mu(R)=R^{-1}\sigma_\mu~ R~,
\end{equation}
with $\mu = 0,\pm 1$
form a $q$-vector in the coproduct algebra of $J_{\mu}$ defined by 
(\ref{JPM}) and (\ref{JZERO}).
A vector is an irreducible tensor of rank $\lambda = 1$. The proof given
below is valid for any $\lambda$.
Let us consider a $q$-tensor $U^{\lambda}_{\mu}$ which
is irreducible in the space generated by $S_\mu$. By definition it must
obey the relations \cite{BID89}:
\begin{equation}\label{WE0}
[S_0,U^{\lambda}_{\mu}] = \mu~U^{\lambda}_{\mu}~,
\end{equation} 
\begin{equation}\label{WE1}
(S_{\pm} U^{\lambda}_{\mu} - q^\mu~U^{\lambda}_{\mu}~S_{\pm})q^{S_0}=
\sqrt{[\lambda\mp\mu][\lambda\pm\mu+1]}~U^{\lambda}_{\mu}~.
\end{equation}
The operator $\sigma_\mu$ defined by (\ref{SIPM}) and (\ref{SIZERO})
is  an example of $U^{\lambda}_{\mu}$ with $\lambda = 1$. 
In the composite system of the coproduct algebra of $J_{\mu}$ a tensor 
$W^{\lambda}_{\mu}$  defined by
\begin{equation}\label{A2} 
W^{\lambda}_{\mu} = R^{-1}~U^{\lambda}_\mu~R~,  
\end{equation}
is irreducible if it satisfies relations analogous to (\ref{WE0})
and (\ref{WE1})
but with $J_\mu$ instead of  $S_\mu$.
Suppose $W^{\lambda}_{\mu}$ satisfies such relations. Below we
show that they are compatible with (\ref{WE0})
and (\ref{WE1}).
  
The validity of 
\begin{equation}\label{WE0P}
[J_0,W^{\lambda}_{\mu}] = \mu~W^{\lambda}_{\mu}~,
\end{equation}
is immediate due to the independence of $J_0$ of $q$, see Eq.(\ref{JZERO}).
Using (\ref{JPM}) the analogue of (\ref{WE1}) becomes
\begin{eqnarray}\label{A3}
&&\left((L_\pm~q^{-S_0}~+~S_\pm~q^{L_0})~R^{-1}~U^{\lambda}_\mu~R-q^\mu~R^{-1}~
U^{\lambda}_\mu~R(L_\pm~q^{-S_0}+S_\pm~q^{L_0})\right)q^{L_0+S_0}\nonumber\\
&&=\sqrt{[\lambda\mp\mu][\lambda\pm\mu+1]}~R^{-1}U^{\lambda}_{\mu\pm1}R~,
\end{eqnarray}
for $W^{\lambda}_{\mu}$ defined by (\ref{A2}).
We multiply the above equation by $R$ on the left and by $R^{-1}$ on
the right and use Eq. (\ref{r8}) to shift the $R$ from the left to the right of 
$L_\pm~q^{-S_0}~+~S_\pm~q^{L_0}$. Using the identity $RR^{-1}=1$
we obtain
\begin{eqnarray}\label{A4}
&&\left((L_\pm~q^{-S_0}~+~S_\pm~q^{L_0})~U^{\lambda}_\mu-q^\mu~
U^{\lambda}_\mu~(L_\pm~q^{-S_0}+S_\pm~q^{L_0})\right)q^{L_0+S_0}\nonumber\\
&&=\sqrt{[\lambda\mp\mu][\lambda\pm\mu+1]}~U^{\lambda}_{\mu\pm1}~.
\end{eqnarray}
Next we use 
\begin{equation}\label{A5}
q^{S_0}~U^{\lambda}_\mu = q^{\mu}~U^{\lambda}_\mu~q^{S_0}~,
\end{equation}
which is a consequence of (\ref{WE0}) and
\begin{equation}\label{A6}
q^{-L_0}~U^{\lambda}_\mu = q^{\mu}~U^{\lambda}_\mu q^{-L_0}~,
\end{equation}
which is a consequence of (\ref{COMM}).
These relations
help to cancel out two of the four terms in the left hand side of (\ref{A4}).
The resulting equation is (\ref{WE1}) which proves that (\ref{A3}) is correct.
Identifying $W^{\lambda}_{\mu}$ with $\Sigma_{\mu}$ i.e. setting $\lambda = 1$
in (\ref{WE0P}) and (\ref{A3}) we obtain equations of type (\ref{r4}) 
for $\Sigma_{\mu}$ i.e. we prove that $\Sigma_{\mu}$ is a $q$-vector in the 
coproduct algebra $J_{\mu}$.

\section{Appendix B}
In this Appendix we show
how the formulae (\ref{EPLUS}) and (\ref{EMINUS}) can be obtained.
For this purpose we need the expectation values of the scalars (\ref{SCALARS}).
In order to calculate explicitly the expectation value of the first and third
scalar products from the list (\ref{SCALARS}) we need the operators 
$\Sigma_{\mu} (\mu = 0,\pm 1)$. which can be obtained by introducing
Eq. (\ref{RQ}) in  Eq. (\ref{SIGQ}). This gives:
\begin{eqnarray}\label{SIGR}
&&\Sigma_1(R)=q^{-2L_0}\sigma_1~,\nonumber\\
&&\Sigma_0(R)=\sigma_0-[2]\lambda\Lambda_{-1}\sigma_1~,\nonumber\\
&&\Sigma_{-1}(R)=q^{2L_0}\sigma_{-1}-[2]\lambda~ q^{L_0}
\Lambda_{-1}q^{L_0}\sigma_0+[2]\lambda^2 q^{L_0}\Lambda^2_{-1}
q^{L_0}\sigma_1~.
\end{eqnarray}
To calculate the expectation value of the second and fourth scalar
products (\ref{SCALARS}) we need
\begin{eqnarray}\label{SIGRMATH}
&&\Sigma_{1}({\mathcal R})=q^{2L_0}\sigma_{1}-[2]\lambda~ q^{L_0}
\Lambda_{1}q^{L_0}\sigma_0+[2]\lambda^2 q^{L_0}\Lambda^2_{1}
q^{L_0}\sigma_{-1}~,\nonumber\\
&&\Sigma_0({\mathcal R})=\sigma_0-[2]\lambda\Lambda_{1}\sigma_{-1}~,\nonumber\\
&&\Sigma_{-1}({\mathcal R})=q^{-2L_0}\sigma_{-1}~,
\end{eqnarray}
which have been derived from the formulae (\ref{ROVERQ}) and (\ref{SIGOVERQ}).

For the purpose of this Appendix, as an example, we first calculate
the expectation value of the third scalar product 
from the list (\ref{SCALARS}). This is
\begin{equation}\label{LAMSIGR}
\vec{\Lambda}\vec{\Sigma}(R)=
-{1\over
q}\Lambda_1\Sigma_{-1}(R)+\Lambda_0\Sigma_0(R)-q\Lambda_{-1}\Sigma_1(R)~.
\end{equation}
From this expression only the first  
and second terms
bring a nonvanishing contribution 
to the expectation value when 
$j = \ell+1/2$.
Looking at the expression of $\Sigma_{-1}(R)$ above we see that 
only the second term contributes so that
$-1/q~\Lambda_1\Sigma_{-1}(R)$ has a nonvanishing contribution due to 
\begin{equation}
\frac{[2]}{q}~\lambda \Lambda_1~q^{L_0}\Lambda_{-1}q^{L_0}\sigma_0.
\end{equation}
Using the definition (\ref{LAPM}) one can rewrite this operator as
\begin{equation}
-\frac{\lambda}{q}~~L_+~L_-\sigma_0.
\end{equation}
At this stage we need the relation 
\begin{equation}\label{47}
L_+~L_-~Y_{\ell m}(q,x_0,\varphi)~=~[\ell+m]~
[\ell-m+1]~Y_{\ell m}(q,x_0,\varphi).
\end{equation} 
For the particular case of $m = \ell$ we have 
\begin{equation}
L_+~L_-~Y_{\ell \ell}(q,x_0,\varphi)~=~[2 \ell]
~Y_{\ell \ell}(q,x_0,\varphi).
\end{equation}
The relation (\ref{47}) has a spin counterpart
\begin{equation}
S_+~S_-~\chi_{m_s}~=~[s+m_s]~
[s-m_s+1]~\chi_{m_s}. 
\end{equation}
Together with (\ref{SIZERO}) this gives
\begin{equation}\label{SIGMA_0}
\Sigma_0~\chi_{1/2} = \sigma_0~\chi_{1/2} = \frac{q}{[2]}~S_+~S_-~\chi_{1/2} =
\frac{q}{[2]}~~\chi_{1/2}.
\end{equation}
Altogether we get
\begin{equation}\label{TERM1}
-\frac{1}{q}~\Lambda_1~\Sigma_{-1}(R)~ \psi_{\ell+1/2,\ell+1/2} =
-\frac{\lambda~[2 \ell]}{[2]}~ \psi_{\ell+1/2,\ell+1/2}.
\end{equation}
According to (\ref{SIGR}) the nonzero contribution of
$\Lambda_0\Sigma_0(R)$ acting on $\psi_{\ell+1/2,\ell+1/2}$ comes from
$\Lambda_0\sigma_0$. Using $\Lambda_0$ as defined by (\ref{LAZERO})
and the relation (\ref{47}) we get
\begin{equation}
\Lambda_0 ~Y_{\ell \ell} = \frac{q}{[2]} [2 \ell]~ Y_{\ell \ell}~.
\end{equation}
Together with (\ref{SIGMA_0}) this gives 
\begin{equation}\label{TERM2}
\Lambda_0 \Sigma_0(R) ~ \psi_{\ell+1/2,\ell+1/2} =
\frac{q^2}{[2]^2}~[2 \ell] ~ \psi_{\ell+1/2,\ell+1/2}~.
\end{equation}
The addition of (\ref{TERM1}) and (\ref{TERM2}) leads to the following
expectation value:
\begin{equation}\label{APLUS1}
{\langle \vec{\Lambda}\vec{\Sigma}(R) \rangle}_{\ell+1/2}
 = {q^{-2}\over[2]^2}[2l]~.
\end{equation}
In the same representation, i. e. $j = \ell+1/2$ the expectation value of 
$\vec{\Sigma}(R)\vec{\Lambda}$ is even easier to obtain inasmuch as
only the term $\Sigma_0(R)\Lambda_0$ contributes. Using the result (\ref{TERM2})
one gets 
\begin{equation}\label{APLUS3}
{\langle \vec{\Sigma}(R)\vec{\Lambda} \rangle}_{\ell+1/2}
={q^2\over[2]^2}[2l]~.
\end{equation}
In a similar manner as above we obtain
\begin{eqnarray}\label{APLUS2}
{\langle \vec{\Sigma}({\mathcal R})\vec{\Lambda} \rangle}_{\ell+1/2}
={q^{-2}\over[2]^2}[2l]~,
\end{eqnarray}
and
\begin{equation}\label{APLUS4}
{\langle \vec{\Lambda}\vec{\Sigma}({\mathcal R}) \rangle}_{\ell+1/2}
={q^2\over[2]^2}[2l]~. 
\end{equation}
By using the expectation values (\ref{APLUS1})-(\ref{APLUS4}),
together with (\ref{RCRUCERP}) and (\ref{RCRUCER}) one 
can calculate the expectation value of (\ref{OPERATOR})
which leads straightforwardly to (\ref{EPLUS}).

For the representation $j=\ell-1/2$, in a similar but longer way  
one obtains
\begin{eqnarray}\label{AMINUS}
{\langle \vec{\Sigma}(R)\vec{\Lambda} \rangle}_{\ell-1/2}
={\langle \vec{\Lambda}\vec{\Sigma}({\mathcal
R}) \rangle}_{\ell-1/2} =-{q^2\over[2]^2}[2l+2]~,\nonumber\\
{\langle \vec{\Sigma}({\mathcal R})
\vec{\Lambda} \rangle}_{\ell-1/2}
={\langle \vec{\Lambda}\vec{\Sigma}(R) \rangle}_{\ell-1/2}
=-{q^{-2}\over[2]^2}[2l+2]~.
\end{eqnarray}
The relations (\ref{RCRUCERP}), (\ref{RCRUCER}) and (\ref{AMINUS})  lead to 
the expectation value (\ref{EMINUS}).

\vspace{2cm}
{\bf Acknowledgements}. We are grateful to David Brink for 
an useful observation and to Daniel Bartz for help with graphics.


\newpage
\listoffigures

\newpage
\begin{figure}
\begin{center}
\psfig{figure=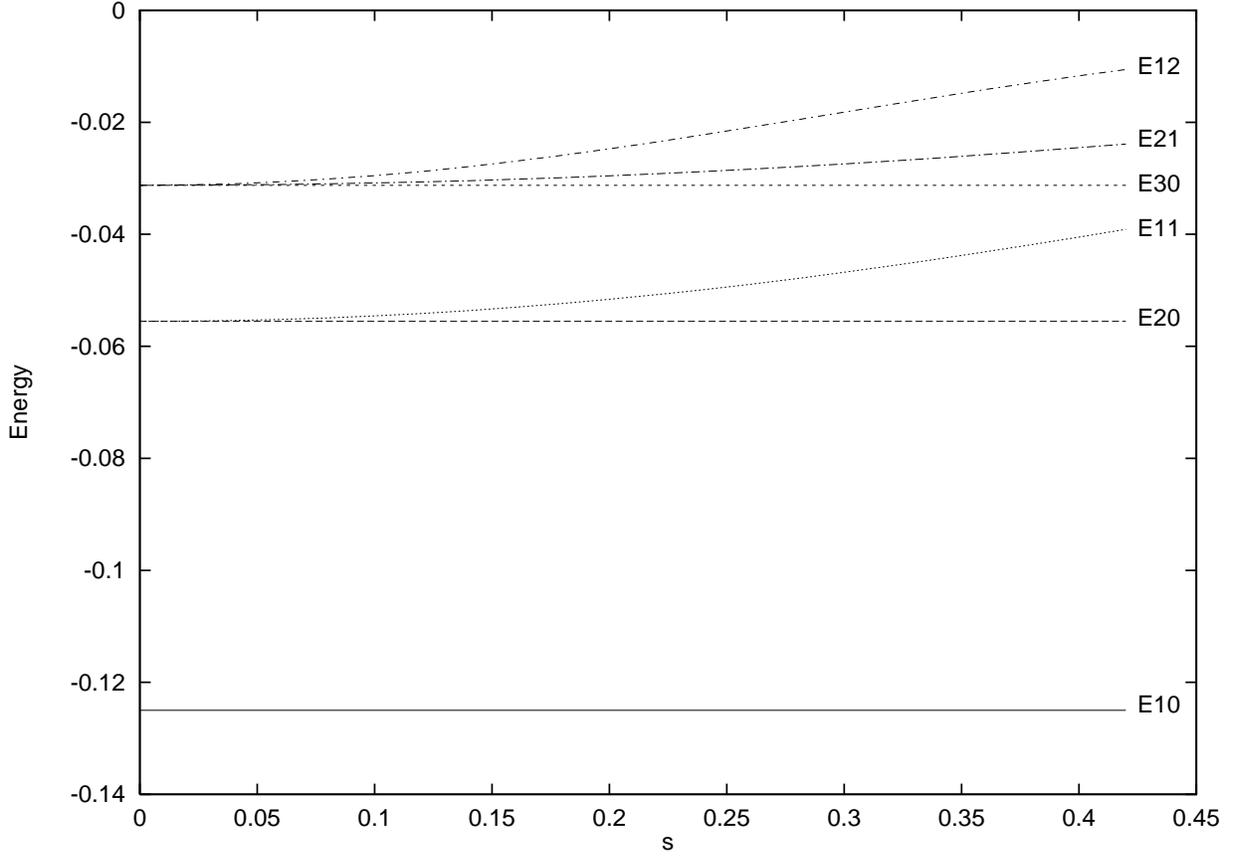,width=16.5cm}
\end{center}
\caption{\label{Fig1}Eigenvalues $\left(E_{n \ell}\right)_{Coulomb}$ 
of Eq. (\ref{COUL})  
as a function of $s$ for a deformation parameter of type (\ref{QREAL}).
The identification with the spectroscopic notation is $E_{10} = E_{1s},
E_{20} = E_{2s}, E_{11} = E_{2p}, E_{30} = E_{3s}, E_{21} = E_{3p}$
and $E_{12} = E_{3d}$.}
\end{figure}

\newpage
\begin{figure}
\begin{center}
\psfig{figure=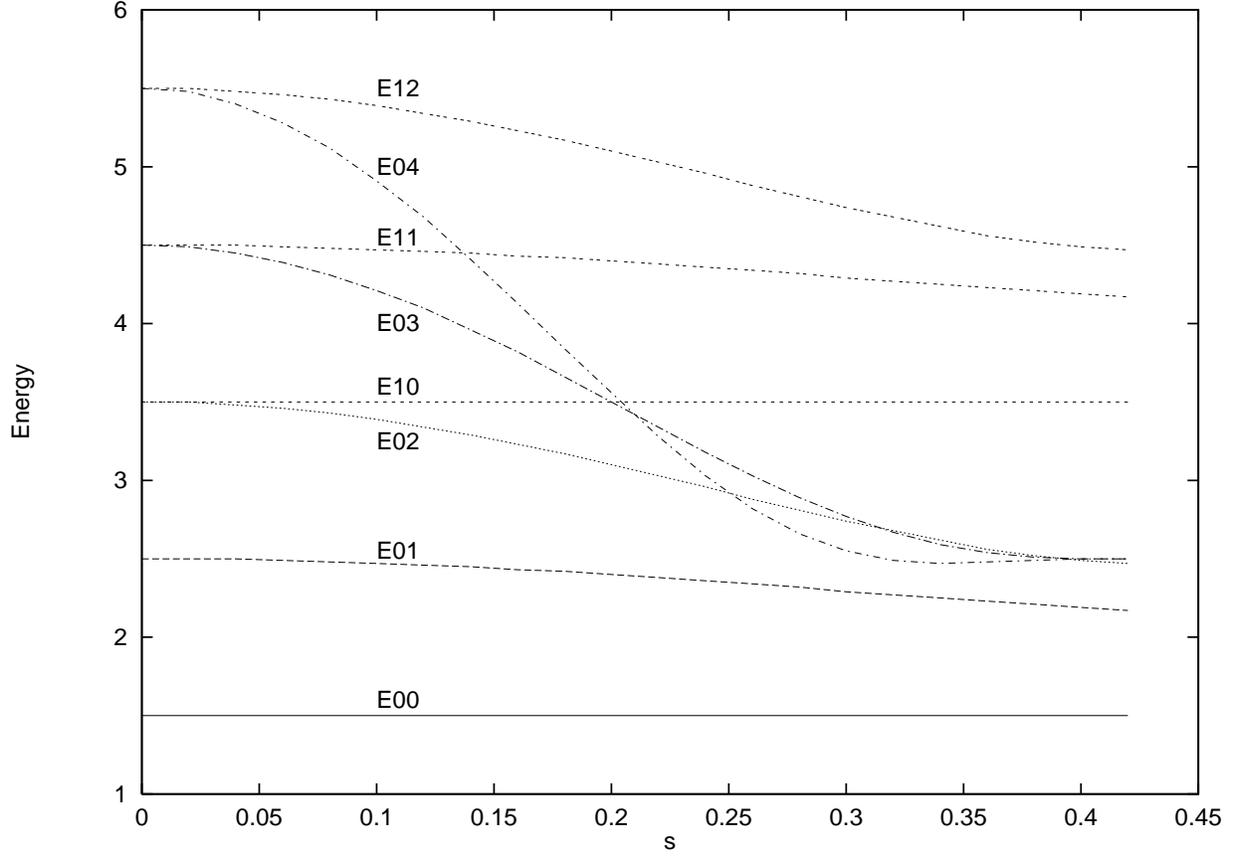,width=16.5cm}
\end{center}
\caption{\label{Fig2}Eigenvalues $(E_{n \ell})_{oscillator}$ of Eq. (\ref{HO}) 
as a function of $s$ for a deformation parameter of type (\ref{QCOM}).
The identification with the spectroscopic notation is $E_{00} = E_{1s},
E_{01} = E_{1p}, E_{02} = E_{1d}, E_{10} = E_{2s}, E_{03} = E_{1f},
E_{11} = E_{2p}, E_{04} = E_{1g}$ and $E_{12} = E_{2d}$.}
\end{figure}

\newpage
\begin{figure}
\begin{center}
\psfig{figure=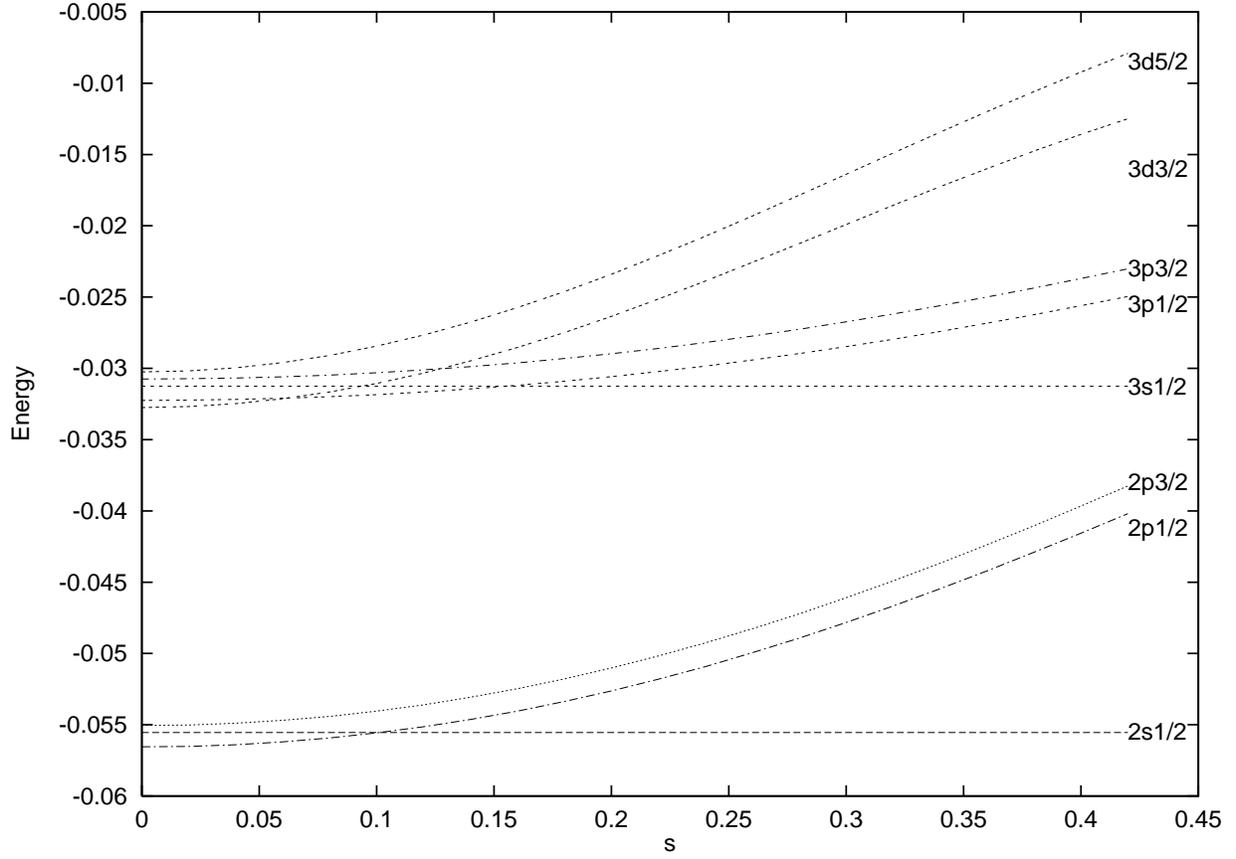,width=16.5cm}
\end{center}
\caption{\label{Fig3}  $\left(E_{n \ell}\right)_{Coulomb}
 + \alpha E_{\ell \pm 1/2}$ with $\alpha$ = 0.001 
as a function  of $s$ for a deformation parameter of type (\ref{QREAL}).}
\end{figure}
 
\newpage
\begin{figure}
\begin{center}
\psfig{figure=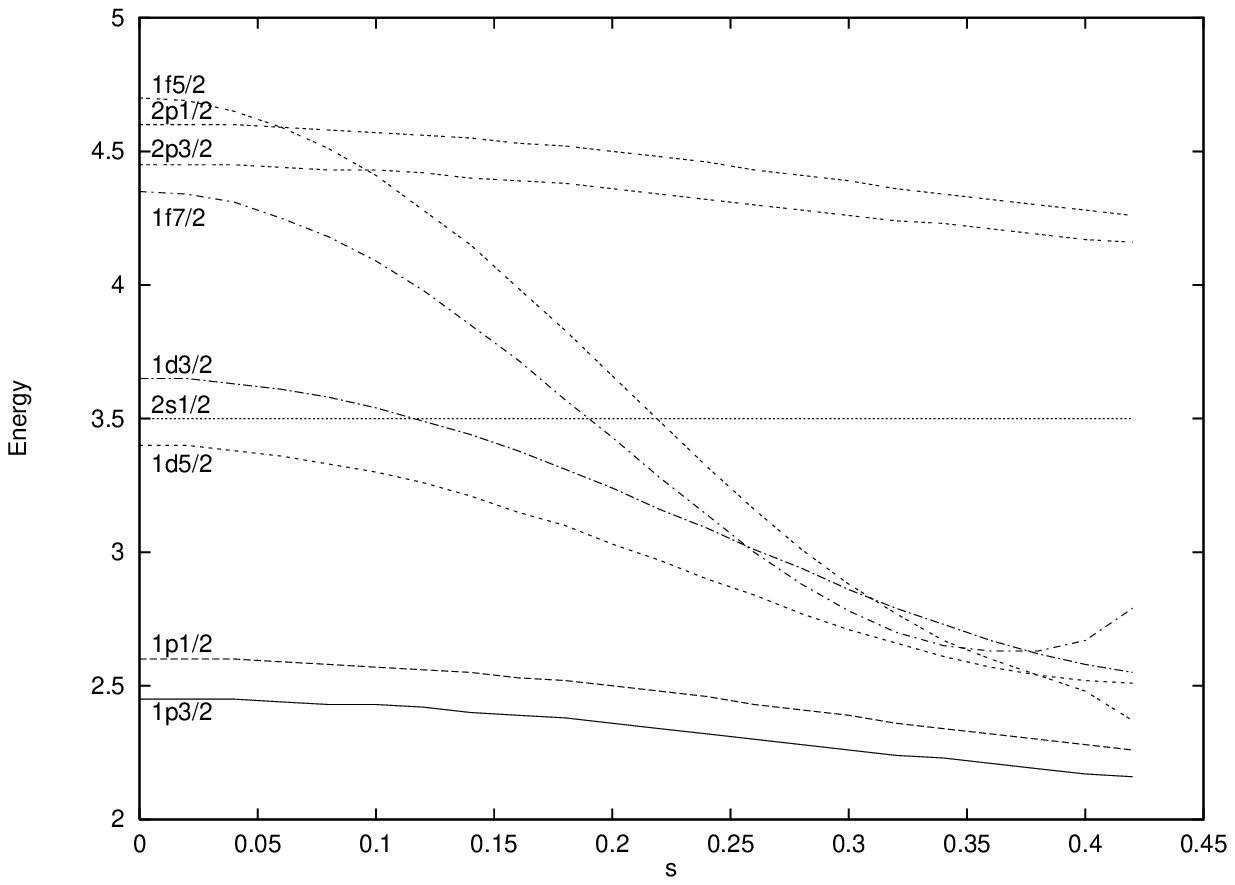,width=16.5cm}
\end{center}
\caption{\label{Fig4} $(E_{n \ell})_{oscillator}
 + \alpha E_{\ell \pm 1/2}$ with $\alpha$ = - 0.1
as a function  of $s$ for a deformation parameter of type (\ref{QCOM}).
The ground state energy $E_{1s1/2} = 1.5$, which is independent of $s$,
is not drawn. }
\end{figure}


\begin{references}
\bibitem{KU81} Kulish P. P. and Reshetikin N. Yu. 1981 Zapiski Semenarov LOMI
{\bf {101}} 101
\bibitem{SK82} Sklyanin E. K. 1982 Funct. Anal. Appl. {\bf {16}} 262
\bibitem{JI86} Jimbo M. 1986 Lett. Math. Phys. {\bf {11}} 247
\bibitem{BI89} Biedenharn L. C. 1989 J. Phys. A: Math. Gen. {\bf {22}} L873
\bibitem{MA89} Macfarlane A. J. 1989 J. Phys. A: Math. Gen. {\bf {22}} 4581
\bibitem{BO91} see e.g. Bonatsos D., Drenska S. B., Raychev P. P., Roussev R. P.
and Smirnov Yu. F. 1991 J. Phys. G: Nucl. Part. Phys. {\bf {17}} L67 and
references therein;
for a recent review on applications of quantum algebras to
nuclear physics see Bonatsos D. and Daskaloyanis C 1999 Progr. Part.
Nucl. Phys. {\bf {43}} 337
\bibitem{MINAHAN} Minahan J. A. 1990 Mod. Phys. Lett. {\bf A5} 2625
\bibitem{LI92} Li You-quan and Sheng Zheng-mao 1992 J. Phys. A: Math. Gen.
{\bf {25}} 6779
\bibitem{CAROW} Carow-Watamura U. and Watamura S. 1994 Int. J. Mod. Phys.
{\bf A9} 3989
\bibitem{XI92} Xing-Chang Song and Li Liao 1992 J. Phys. A: Math. Gen.
{\bf {25}} 623
\bibitem{IR66} Irac-Astaud M. 1996 Lett. Math. Phys. {\bf {36}} 169
\bibitem{PAPP} Papp E. 1996 J. Phys. A: Math. Gen. {\bf {29}} 1795 
\bibitem{MI99} M. Micu 1999 J. Phys. A: Math. Gen. {\bf {32}} 7765
\bibitem{RAYCHEV98} Raychev P. P., Rousev R. P., Lo Iudice N. and
Terziev P.A., 1998 J. Phys. G: Nucl. Part. Phys. {\bf {24}} 1931
\bibitem{RAYCHEV96} Raychev P. P., Rousev R. P., Terziev P.A.
Bonatsos D. and  Lo Iudice N. 1996 J. Phys. A: Math. Gen. {\bf {29}} 6939
\bibitem{ANDREWS} Andrews G., Askey R. and Roy R. {\it Special functions},
Cambridge University Press, 1999 
\bibitem{CU91} Curtright T. L., Ghandour G. I. and Zachos C. K.
1991 J. Math. Phys. {\bf {32}} 676
\bibitem{RS92} Rittenberg V. and Scheunert M. 1992, J. Math. Phys. {\bf 33} 436
\bibitem{QUESNE} Quesne C. 1993 J. Phys. A: Math. Gen. {\bf {26}} L299 
\bibitem{GR91} Grosse H., Martin A. and Stubbe J. 1991 Phys. Lett.
{\bf {B255}} 563
\bibitem{GR94} Grosse H., Martin A. and Stubbe J. 1994 J. Math. Phys.
{\bf {35}} 3805
\bibitem{LU97} see e.g. Lucha W and Sch\"oberl F. F, 1997 Phys. Rev. {\bf A56} 139,
Martin A. 1997 {\it Quark Confinement and the Hadron Spectrum II} eds. Brambilla N.
and Prosperi G. M. World Scientific, Singapore p.187
\bibitem{MA} Martin A, talk given at the
workshop {\it Critical Stability of Quantum Few-Body Systems},
ECT* Trento, February 3-14 1997, unpublished
\bibitem{BO69} Bohr A. and Mottelson B. R. 1969 {\it Nuclear Structure},
Benjamin New York, vol. 1, chapter 2, sec. 4
 
\bibitem{GR92} Grosse H., Martin A. and Stubbe J. 1992 Phys. Lett.
{\bf {B284}} 347
\bibitem{BR70} Brink D. M. and Vautherin D. 1970 Phys. Lett.
{\bf {32B}} 149
\bibitem{BID89}  Biedenharn L. C. and Tarlini M. 1990 ~Lett. Math. Phys.
{\bf 20} 271 
\end{references}
\end{document}